\documentstyle[aps,multicol,epsf,12pt]{revtex}

\begin{document}

\title{Spin-engineered quantum dots}
\author{
V. Fleurov$^{1}$, V.A. Ivanov$^{2,3}$, F.M. Peeters$^2$, and
I.D. Vagner$^{4,5}$ \\ 
$^1$Beverly and Raymond Sackler Faculty of Exact Sciences, School of Physics
and Astronomy, Tel Aviv University Tel, Aviv 69978, Israel.\\
$^2$Departement Natuurkunde, Universiteit Antwerpen (UIA), Universiteitplein
1, B-2610, Antwerpen, Belgium\\
$^3$N.S. Kurnakov Institute of General and Inorganic Chemistry of the
Russian Academy of Sciences, Leninskii prospect 31, 117907, Moscow, Russia\\
$^4$Grenoble High Magnetic Field Laboratory (MPI/FKF \& CNRS),
BP166X, F-38042, Grenoble Cedex 9, France\\
$^5$Department of Communication Engineering, Holon Academic
Institute of Technology, POB 305, Holon 58102 Israel.} \maketitle

\begin{abstract}
Spatially nonhomogeneously spin polarized nuclei are proposed as a new
mechanism to monitor electron states in a nanostructure, or as a means to
createn and, if necessary, reshape such nanostructures in the course of the
experiment. We found that a polarization of nulear spins may lift the spin
polarization of the electron states in a nanostructure and, if sufficiently
strong, leads to a polarization of the electron spins. Polarized nuclear
spins may form an energy landscape capable of binding electrons with energy up
to several meV and the localization radius $ >$ 100\AA. 
\end{abstract}

PACS numbers; 85.30.V, 31.30.G, 33.35

\section{Introduction}

Progress in microelectronics depends crucially on a deep understanding of the
electronic properties of low dimensional semiconductors and
nanostructures. Typical examples are heterojunctions, quantum wells, and
fabricated out of them, 1D quantum wires and 0D quantum dots.

Characteristics of nanostructure devices are mainly determined by the shape of
the potential landscape and can be monitored by applying a gate voltage and/or
an external magnetic field. A possible role of polarized nuclear spins is
generally overlooked in this context. The technique of interband optical
pumping, developed in the 60th \cite{l68} (see also \cite{plss77}) opened up a
way to reach nuclear spin polarization approaching 50\%. This results in an
effective magnetic field (called Overhauser field) acting on the electron
spins which may reach a few Tesla, or equivalently, several meV energy. This
energy is comparable to the electron energies in typical nanostructure devices.

As a consequence, one may expect that through a controlled polarization or
depolarization of nuclear spins one may monitor the characteristics of a
nanodevice. The Overhauser field leads to an effective Zeeman splitting of
the electron states and subsequently to their shifts. If this effect is
sufficiently strong, it may lead to a polarization of the electrons in the
dot.

Since, at low temperature, the characteristic time of nuclear spin relaxation
is extremely long, one may think about creating a potential landscape for the
electrons by a spatially inhomogeneous polarization of the nuclei. By
polarizing nuclei in a small region one can create a local potential,
attractive for electrons with one spin orientation and repulsive for electrons
with the opposite spin orientation. This may open up a new way of
spin-engineering of quantum dots or geometrically more complex nanodevices,
whose shapes can be manipulated in real time.

All semiconductor materials consist of more than one stable elemental isotope
with non zero nuclear spin $I$; for example $^{69}$Ga (natural abundance
60.4\%), $^{71}$Ga (31.6\%), $^{75}$As (100\%), all have nuclear spin $I =
\frac{3}{2}$, while $^{27}$Al has $I = \frac{5}{2}$. A lithographically
prepared GaAs/Al$_{1-x}$Ga$_x$As heterostructure of dimensions
10nm$\times$10nm comprises a huge number of active nuclear spins $\sim
10^4$. In (Al)GaAs the amount of active nuclei is of the order of the total
number of atoms in that volume. In  Si/Si$_{1-x}$Ge$_x$ heterojunctions the
active nuclear spins have substantially lower concentration because the
natural abundances of the Ge and Si isotopes with nonzero nuclear spins are
rather small (7.6\% for $^{73}$Ge with $I = \frac{9}{2}$ and 4.7\% for
$^{29}$Si with $I = \frac{1}{2}$). However, recent progress in nanotechnology
and growth of isotopically controlled bulk Si \cite{tioso99} and Ge
\cite{mimmush00}, superlattices of, e.g., $^{70}$Ge/$^{74}$Ge \cite{ih01} will
allow one to fabricate low dimensional semiconductor structures with
controlled abundances of  spin active and neutron transmuted nuclei.

Nuclear spin diffusion decreases with the dimensionality \cite{mbh01} of the
system. Earlier estimates \cite{pbs57} show that nuclear spin relaxation times
in conventional pure semiconductors are very long. At helium temperature they
are at least of the order 10$^2$ to 10$^3$ s and can reach up to a few
hours. As a comparison, the electron spin relaxation times are about
10$^{-7}$s. It means that any time variation of the nuclear polarization
occurs adiabatically slow as compared to the time scales of the electron
dynamics. Therefore, one may consider any potential created by the polarized
nuclei for the electron subsystem as quasi-static. Vagner etal \cite{vrwz98}
proposed a new type of Aharonov - Bohm effect caused by spin polarized nuclei
in the absence of any magnetic field, which was observed very recently
\cite{bbgikmrs01}. Ref. \cite{bfv98} discusses an anomalous Hall effect caused
by the hyperfine interaction of polarized nuclei and electrons. In a recent
paper \cite{enf01} it was shown that a rather small nuclear spin polarization
can contribute to the electron relaxation in a quantum dot.

Due to the enormous difference of nuclear, $m_n$, and electron, $m$, masses,
the Zeeman splitting is substantially smaller for nuclei as compared to that
for electrons. It makes polarization of nuclei by an external magnetic field
much more difficult. Nevertheless, various optical techniques
\cite{l68,plss77,ka00} for polarizing nuclear spins via creating
nonequilibrium spin polarized electrons, which transfer their polarization to
the nuclear subsystem in the course of thermal equilibration of electrons,
lead to much better results. These techniques are much more efficient than the
direct magnetization by an external magnetic field and result in much higher
nuclear spin polarizations.

Overhauser \cite{o53,o54} has described the hyperfine interaction of
the electron and the nuclear spins in a solid through the Fermi-like contact
potential
\begin{equation}\label{1}
H_{hf} = A{\bf S}\sum_i {\bf I_i}\delta({\bf r} - {\bf r_i}) \equiv g\mu_B
\hat {\bf B}_n\cdot{\bf S} 
\end{equation}
where the summation in (\ref{1}) is over the spin active nuclei, $\mu_B$ is
the Bohr magneton, and $g$ is the electron $g$-factor. $A$ is the coefficient
of the hyperfine interaction, which is specific for each particular type of
nucleus.  The operator $\hat{\bf B}_n$  averaged over the nuclei and the
electon wave functions results in the hyperfine Overhauser field, ${\bf B}_n$,
which acts on the electron spin. In GaAs the hyperfine coefficient $A$ is
negative and the Overhauser field tends to polarize the electron spins
parallel to the nuclear spins. This field may be large
\cite{s90,dp84,bgk90,wkm94,bdpwt95} depending on the type of nuclei and the
degree of their polarization. For example, for naturally abundaned isotopes in
GaAs it reaches the value $B_n = 5.3$T in the limit that all nuclear spins are
completely polarized. The value $B_n =1.7$T has been achieved experimentally
\cite{plss77}, which corresponds to 32\% nuclear polarization. Thus electrons
may be strongly influenced by the Overhauser field in a nano-device with
polarized nuclear spins, although this field does not manifest itself
magnetically due to the smallness of the nuclear magnetic moments.

\section{Quantum dot with polarized nuclear spins.}

The aim of this section is to estimate the influence of the Overhauser field,
created by a spatially nonhomogeneous nuclear polarization, on the electrons
in a quantum dot. We assume that a local Gaussian distribution of nuclear
polarization along the $z$ axis was created,
\begin{equation}\label{2}
I_i = I_m\exp\left(-\frac{\rho_i^2}{2a^2} \right) \exp \left(-
\frac{z_i^2}{2b^2} \right),
\end{equation}
where $\{\rho_i,z_i\}$ are the cylindrical coordinate of the $i$-th
nucleus. A quantum dot (QD) with spin polarized nuclei is described by the
Hamiltonian
\begin{equation}\label{3}
H = \left(-\frac{\hbar^2}{2m}\Delta + \frac{m\omega^2r^2}{2}\right)
\delta_{\sigma,\sigma'} + H_{hf},
\end{equation}
where for simplicity we assumed a 3D parabolic confined quantum dot with
confinement frequency $\omega$. The QD part in (\ref{3}) is diagonal with
respect to the spin projections $\sigma$ whereas $H_{hf}$ is defined by
Eq. (\ref{1}) with the nuclear polarization distribution (\ref{2}). Here we
shall disregard the spin flip processes and consider only the longitudinal
part of the hyperfine interaction, i.e., we assume that ${\bf S}\cdot {\bf
I_i} = I_i\sigma_z$.

The eigenenergies and eigenfunctions of the three dimensional harmonic
oscillator representing the quantum dot in the absense of nuclear spin
polarization is well known:
\begin{equation}\label{4}
E_{n_x,n_y,n_z} = \hbar\omega \left(n_x + n_y + n_z + \frac{3}{2}\right),
\end{equation}
\begin{equation}\label{5}
\Psi_{n_x,n_y,n_z}({\bf r}) = \frac{\xi^{3/2}}{(2^{n_x + n_y +
n_z}n_x!n_y!n_z! \sqrt{\pi})^{1/2}} \exp(-\frac{\xi^2(\rho^2 + z^2)}{2})
H_{n_x}(x\xi)) H_{n_y}(y\xi) H_{n_z}(z\xi)
\end{equation}
where $H_n(...)$ is the Hermite polynomials, and $\xi = \sqrt{m\omega/\hbar}$
determines the inverse size of the dot.

Now we calculate the first order perturbation correction to the ground state
energy
\begin{equation}\label{6}
E^{(1)}_{0\sigma} = \frac{\sigma A}{\Omega} I_m \frac{2\sqrt{2}a^2b\xi^3}{[1 +
2a^2\xi^2] \sqrt{1 + 2 b^2 \xi^2}} 
\end{equation}
where $\Omega$ is the volume per nuclear spin, and $\sigma = \pm 1$ is the
direction of the electron spin. 

The second order correction is due to virtual excitations to even states
of the oscillator for the chosen spatial distribution of the nuclear
spin polarization. The contribution of the lowest 3-fold degenerate state with
the energy $E_{200} = E_{020} = E_{020} = \frac{7}{2}\hbar\omega$ is
\begin{equation} \label{7}
E_0^{(2)} = -\frac{2A^2I_m^2}{\hbar\omega\Omega^2} \frac{a^4b^2\xi^6}{[1 +
2a^2\xi^2]^2 [1 + 2 b^2 \xi^2]}\left\{\frac{2}{[1 + 2 a^2 \xi^2]^2} +
\frac{1}{[1 + 2 b^2 \xi^2]^2}\right\}.
\end{equation}
The energy of the ground state second order perturbation theory becomes
\begin{equation}\label{8}
E_{0\sigma} = \frac{3}{2} \hbar \omega + E^{(1)}_{0\sigma} + E_0^{(2)}. 
\end{equation}
If the nuclei are polarized in a region, exceeding the size of the dot (i.e.,
$a\xi,b\xi > 1$) then this energy becomes
\begin{equation}\label{9}
E_{0\sigma} = \frac{3}{2} \hbar \omega + \sigma\frac{AI_m}{\Omega} -
\frac{A^2I_m^2}{16\Omega^2\hbar\omega} \frac{1}{\xi^4} \left(\frac{2}{a^4} +
\frac{1}{b^4} \right) 
\end{equation}

We see that that the principal effect of the Overhauser field on the electron
levels is the Zeeman-like lift of the spin degeneracy. It holds also for a
homogeneous polarization of the nuclear spins, when $a,b\to \infty$. A
nonhomogeneous polarization may shift the electron levels. The ground state
(\ref{9}) is shifted downward regardless of the $\sigma$ value. However, one
should keep in mind, that accounting for the nuclear spin contribution, the
potential at the bottom of the well becomes now negative, $- \frac{AI_m}{2\Omega}$, meaning that the distance of
the ground state from the bottom of the well may, in fact, increase.

According to the estimates, presented above, the Zeeman-like splitting caused
by the first order correction (\ref{6}) linearly depends on the nuclear
spin polarization and may become substantial. The condition
$\frac{AI_m}{\Omega} > \hbar\omega$ can be easily achieved. Although under
this condition one cannot restrict oneself to the lowest order terms of
perturbation theory, nevertheles the general pattern is qualitatively clear,
which is illustrated in Fig.1. As an example we consider a quantum dot
occupied by six electrons. It is well known that the Pauli principle makes the
total spin of a dot with an even number of electrons equal to zero, when
the nuclear polarization is zero and the spin degeneracy is not lifted, as is
situation shown in Fig. 1(a). At relatively small $\frac{AI_m}{\Omega} <
\hbar\omega$ polarization the spin degeneracy is lifted but the spin
configuration of the electrons in the dot does not change, Fig. 1(b). There
are still three spin up electrons and three spin down electrons, so that the
net spin of the dot remains zero. However, at larger nuclear polarizations,
$\frac{AI_m}{\Omega} > \hbar\omega$, the Zeeman-like splitting becomes larger
than the energy distance between the levels, which results in a new spin
configuration of the electrons in the dot. One can see from Fig. 1(c) that the
level $n=3$ for spin up electrons moves below the  $n=2$ level for the spin
down electrons. As a result one electron flips its spin. We now have four spin
up electrons and only two spin down electrons, which results in a net electron
spin of the dot equal to 1.
\vspace{1cm}

\begin{figure}[htb]
\epsfysize=10 \baselineskip
\centerline{\hbox{\epsffile{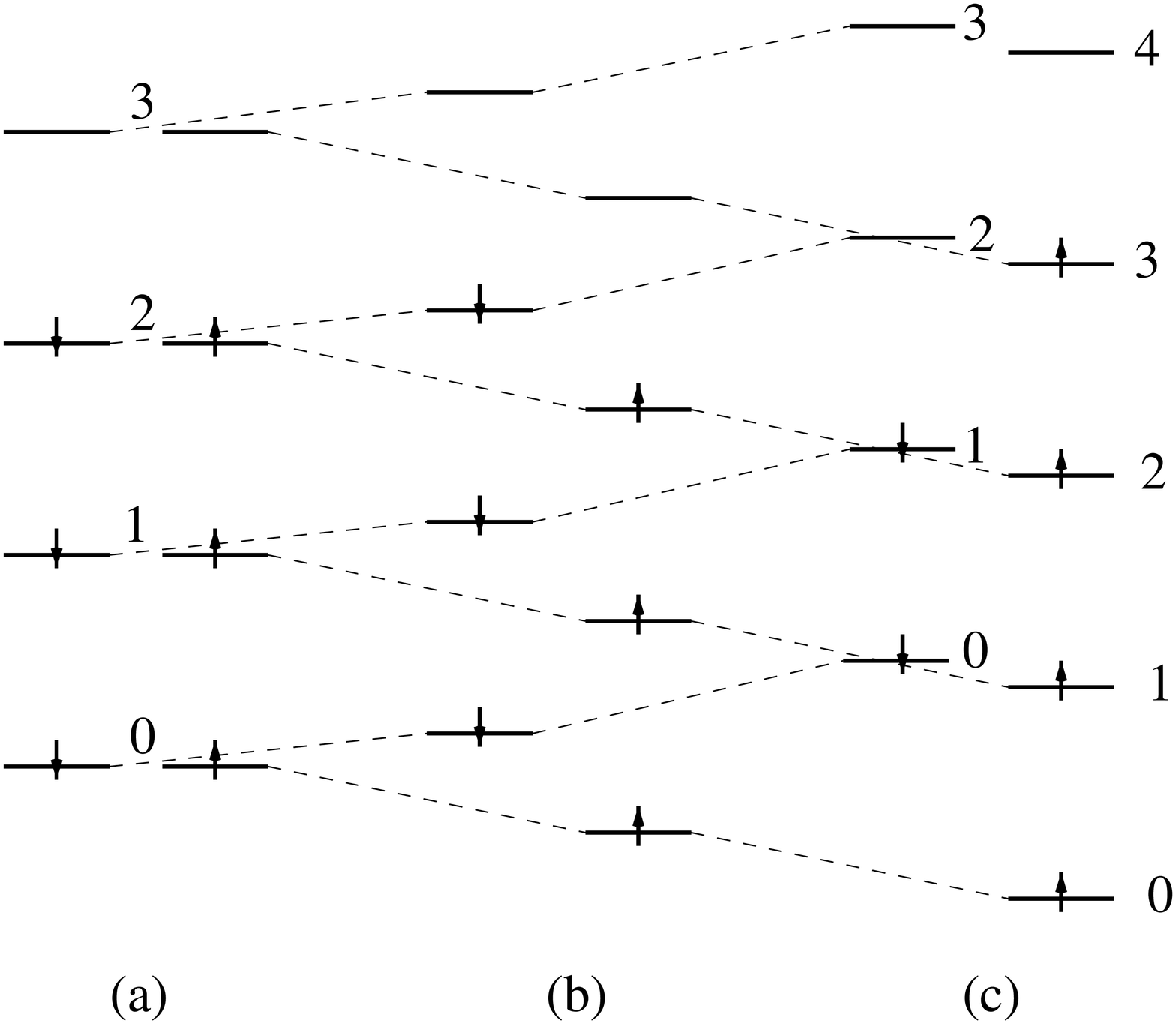}}}{\vspace{1cm}}
\caption{A graphical illustration of the polarization of a quantum dot with
even number of electrons. {\em (a)} shows four lowest levels of the dot
occupied by six electrons in the absence of the nuclear spin
polarization. {\em (b)} The nuclear spin polarization causes a Zeeman-like
splitting of the levels which is still less than the distance between the
levels. No net spin appears in the dot. (c) The Zeeman-like splitting
exceeds the distance between the levels resulting in a spin polarized electron
state.} 
\end{figure}

\section{Potential created by polarized nuclear spins}

Having considered the role played by spin polarized nuclei in a conventional
quantum dot, we address now the problem of engineering potential landscapes
by means of polarized nuclei. This may be achieved by creating a spatially
inhomogeneous nculear spin polarization. For $A < 0$ the resulting potential
(\ref{1}) is attractive for spin up electrons (parallel to the nuclear spin
polarization) and repulsive for the spin down electrons (antiparallel to the
nuclear spin polarzation). One may create a nuclear spin polarization $I_m$ in
a region with a size $a$, thus forming an attractive (for spin up electrons)
potential $U_{hf}= \frac{AI_m}{\Omega}$. Then a simple reasoning based the
uncertaintly principle leads us to the conclusion such a potential may bind an electron if the condition
\begin{equation}\label{10}
a > \hbar\sqrt{\frac{\Omega}{2m^*|A|I_m}} 
\end{equation}
is fulfilled. 

An estimate can be made for the case of GaAs. Creating a 30\% nuclear spin
polarization results in a potential $U_{hf} \sim 3$meV which is capable of
binding an electron in a well of typical size $a > 100$\AA. 

\section{Conclusions}

We discussed the possible influence of the hyperfine interaction between
spin polarized nuclei and electrons and its influence on the electron
subsystem with a special emphasis to confined regions in semiconductors, e.g.,
quantum dots. An effective Overhauser magnetic field of the polarized nuclei
lifts the spin degeneracy and is therefore, capable of polarizing electrons in
a quantum dot. Long lived nonhomogeneously spin polarized nuclei may create a
local potential, attractive for electrons with one spin direction and
repulsive for the electrons with the opposite spin directions. Such a
potential can form, e.g., a quantum well, a quantum dot or any other
nanostructure, depending on the spatial engineering of the polarized
nuclei. The interesting feature of such potential landscapes is that they can
be created and reshaped, if necessary, in real time by polarizing and/or
depolarizing nuclear spins. We believe that this sort of technique, being
developed experimentally, may open a new promising venue in
nanotechnologywhich we call spin - engineering.

{\bf Acknowledgement} The authors are indebted to Max Planck Institute
for Complex Systems, Dresden, for hospitality. VAI and FMP are supported by
the Flemish Science Foundation (FWO-Vl), the Belgium Interuniversity
Attraction Poles (IUAP) and the Flemish Concerted Action (EOA) programme.

\end{document}